# Three terminal capacitance technique for magnetostriction and thermal expansion measurements

B. Kundys [‡], Yu. Bukhantsev, and S. Vasiliev
*Institute of Physics, Polish Academy of Sciences, Al. Lotników 32/46, 02-668, Warsaw, Poland*
D. Kundys
*Department of Physics and Astronomy, University of Sheffield, Sheffield S3 7RH, United Kingdom*
M. Berkowski and V. P. Dyakonov
*Institute of Physics, Polish Academy of Sciences, Al. Lotników 32/46, 02-668, Warsaw, Poland*

An instrument has been constructed to measure a large range of magnetostriction and thermal expansion between room temperature and 4 K in a superconductive split-coil magnet, that allows investigation in magnetic fields up to 12 T. The very small bulk samples (up to 1 mm in size) as well as big ones (up to 13 mm) of the irregular form can be measured. The possibility of magnetostriction investigation in thin films is shown. A general account is given of both electrical and the mechanical aspects of the design of capacitance cell and their associated electronic circuitry. A simple lever device is proposed to increase the sensitivity twice. The resulting obtained sensitivity can be 0.5 Å. The performance of the technique is illustrated by some preliminary measurements of the magnetostriction of superconducting $MgB_2$, thermal expansion of $(La_{0.8}Ba_{0.2})_{0.93}MnO_3$ single crystal and magnetoelastic behavior of the Ni/Si(111) and $La_{0.7}Sr_{0.3}CoO_3$ /$SAT_{0.7}CAT_{0.1}LA_{0.2}$(001) cantilevers.          [DOI: 10.1063/1.1753088]

## I. INTRODUCTION

There are two different reasons to make investigations in magnetostriction. The first one is that in transducer applications we would like to get as much magnetostriction as possible, and some alloy compositions reveal magnetostriction coefficients as high as[1] $10^{-3}$. In this case, as in the case of thermal expansion, the required sensitivity of the measurements is not so extreme and can be measured by a number of other experimental techniques.[2] Most of the alternative methods of measurements however involve large samples, which could not be obtained so easily by existing crystal growth techniques. The second reason for investigation is to keep magnetostriction very low. This is what happens in giant magnetoresistance (GMR) development[3–6], where magnetostriction is not welcomed due to its direct connection with magnetic anisotropy and with magnetic softness of the materials. One more reason to keep magnetostriction low is to minimize the parasitic signals caused by ultrasonic disturbances at the head/media interface, which the magnetostriction would transform into magnetic signals, which the head will further convert into electrical noise, thus decreasing the safety margins of the bit detectors, and increasing the error/read rate. All these head performance parameters are critical to the acceptance. The quality control of the GMR applications often requires the magnetostriction coefficient to be as small as $10^{-7}$ or even less, and the instrumentation checking this must be extremely sensitive. That is why the main attention in this work will be focused on the magnetostriction measurements.

The history of the capacitance dilatometer starts from thermal expansion measurements[7] and up to now a lot of efforts have been made (see Ref. 8 and references therein) in order to improve its advantages. Moreover it is desirable to have at least one well-established technique for the wide range magnetostriction investigation in bulk samples as well as in thin film ones[9,10] and capacitance instrument offers a very good way to accomplish this. In this article we attend to some salient points in the mechanical design of the apparatus, combining our experience with already reported methods, where the aim is to realize the potential advantages that capacitance methods possess in relatively simple construction, high sensitivity and long term stability. This is an adaptation of parallel plate capacitance technique,[11] which allows magnetostriction measurements in the both parallel and perpendicular mode in field up to 12 T in temperature range of 4–300 K. A simple way of increasing sensitivity is shown. The detailed consideration of technical aspects is presented.

## II. EXPERIMENTAL SETUP

### A. Cell description

The construction of the capacitance cell illustrated in Fig. 1 follows closely Tsuya's[11] design [Fig. 1(a)]. The length and diameter of the cell cylinder are about, 28 and 90mm respectively. The cell consists of a fixed electrode (1) and movable one (2), which can move due to magnetostriction or thermal expansion. Both electrodes are isolated from the rest of the cell, using block of PTFE® isolating material (4). The sample (3) is mounted on a copper rod with a differential screw (5).

---

[‡] Electronic mail: kundys@.gmail.com (Bohdan Kundys).

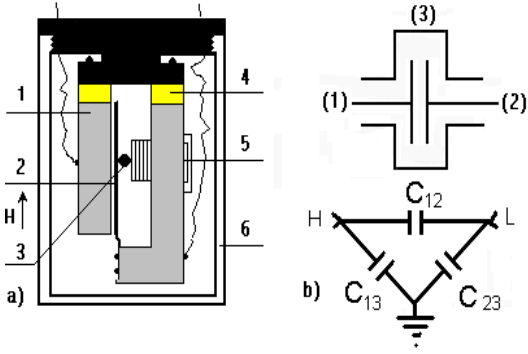

FIG. 1. Schematic of capacitance cell follows Tsuya - Ref. 11 (a), and associated electronic circuitry - Ref. 12 (b) . 1-fixed electrode, 2-movable electrode, 3-sample, 4-rod with differential screw, 5-block of isolating material, 6-shielding case.

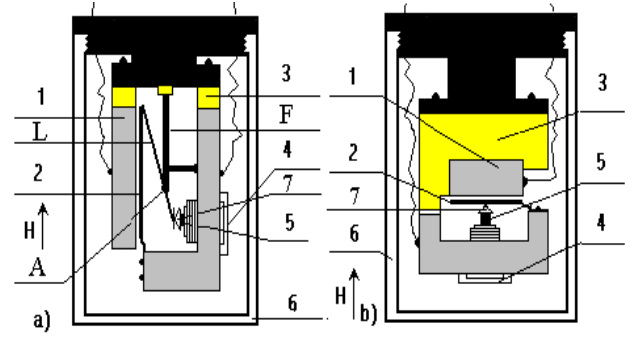

FIG. 2. Capacitance cell with lever device, which allows sensitivity to be increased two times (a). The capacitance cell adapted for longitudinal magnetostriction measurements (b). 1-fixed electrode, 2-movable-electrode, 3-block of isolating material, 4-copper rod with differential screw, 5-sample, 6-shielding case, 7-beryllium copper cap. L-lever, A-fixed point, F-holder.

The movable electrode (2) can change its position in both directions depending on the changes in length of a specimen. The whole cell is surface grounded using copper shielding case, which also secures a good thermal stability (6). The main changeable capacity is between fixed (1) and movable (2) electrodes. The upper side of Fig. 1(b) schematically shows three terminals for the capacitance cell[12]. The lower side of Fig. 1(b) shows the equivalent circuit for the three-terminal approach. [The direct capacitance $C_{12}$ (by movable and fixed electrode), which appears between the HIGH and LOW terminals is combined with whatever undesired stray capacitance may exist. The capacitors $C_{13}$ and $C_{23}$ represent the capacitance of the unknown capacitor plates to surrounding objects (such as the capacitor case and ground). The three-terminal measurement configuration is changed to a two-terminal measurement configuration by connecting the LOW terminal to ground. This eliminates $C_{23}$ and puts $C_{12}$ and $C_{13}$ in parallel so that two terminal measurements cannot separate the two capacitances. $C_{12}$ contains stray capacitance caused by surrounding objects and also capacitance contributions from the coaxial cables that are proportional to the cable length. Therefore, $C_{12}$ cannot be measured accurately, unless it is much larger than $C_{13}$. This is a very serious limitation to precise measurement. Each of (1) and (2) electrodes also make capacity with a shielding case independently [Fig. 1(b) lower side]. That permits a stray capacitance $C_0$ [Eq. (1)] to be fixed, thus making it dependent mainly on the geometry of the cell. The body of the cell is a cylinder of oxygen-free high conductivity copper, which after machining is annealed for 24h in an inert atmosphere at about 500 °C to avoid creep and hysteresis on thermal cycling. The cell is small enough to be introduced into a variable temperature helium flow cryostat with the superconductive coils, providing a variable magnetic field ($\pm$ 12 T).

### B. Bulk samples

When electrodes of a cell are parallel the capacitance of the condenser is:

$$C = \frac{\varepsilon \varepsilon_0 S}{d} + C_0 \quad (1)$$

where $\varepsilon_0$ and $\varepsilon$ are permittivity of free space and dielectric constant of the environment respectively, S -area of the electrodes, d- distance between them. $C_0$ is a stray capacity, which doesn't depend on a distance between the electrodes.

The displacement "$\Delta L$" of movable electrode caused by magnetostriction of the sample leads to the change in capacity:

$$C_1 = \frac{\epsilon\epsilon_0 S}{d+\Delta L} + C_0, \quad (2)$$

$$\Delta L = \epsilon\epsilon_0 S \left[\frac{1}{(C_1 - C_0)} - \frac{1}{(C - C_0)}\right]$$

$$= \epsilon\epsilon_0 S \frac{C - C_1}{C_1(C - C_0) - C_0 C + C_0^2}$$

$$= \epsilon\epsilon_0 S \frac{\Delta C}{C^2 - C_0 C - \Delta C C + \Delta C C_0 - C_0 C + C_0^2}$$

$$= \epsilon\epsilon_0 S \frac{\Delta C}{(C - C_0)^2 - (C - C_0)\Delta C}. \quad (3)$$

Inasmuch as $C_1 = C - \Delta C, and (C - C_0)\Delta C << (C - C_0)^2$

the displacement is equal:

$$\Delta L = \frac{\varepsilon\varepsilon_0 S \Delta C}{(C - C_0)^2} \quad (4)$$

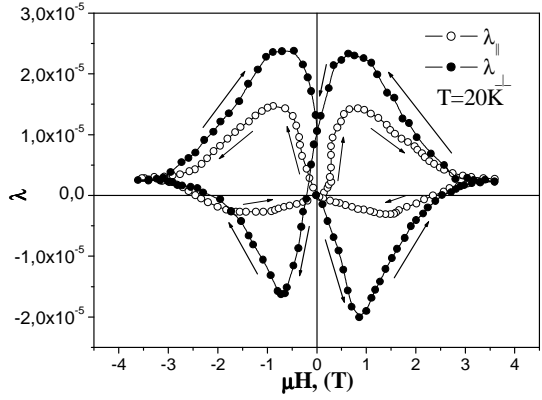

FIG. 3. Magnetostriction loops of MgB2 superconductor taken at 20 K with magnetic field directed perpendicular (l⊥) and parallel (l∥) to the direction of measured magnetostriction.

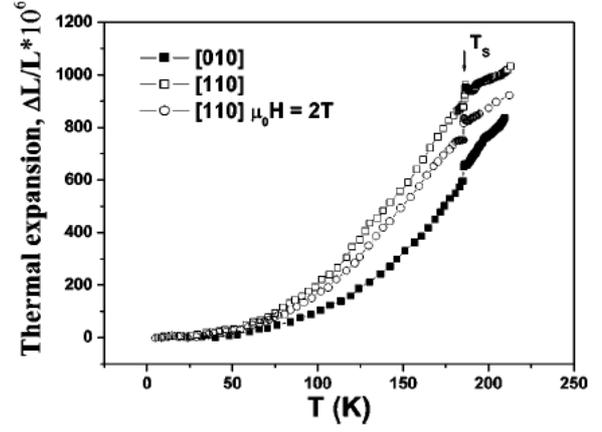

FIG. 4. Thermal expansion of $(La_{0.8}Ba_{0.2})_{0.93}MnO_3$ single crystal measured along specific crystallographic directions.

Figure 2(a) presents our adaptation of Tsuya's capacitance cell with a simple lever device, which allows the increasing of sensitivity. In our design the gain in the sensitivity was obtained to be double. The sample pushes on the one end of the lever, which is reliably fixed at point A using holder F. Another end of the lever is connected to free end of movable electrode (2). The length of the lever rod from fixed point A to the free end of movable electrode is two times bigger than its length from fixed point to the sample. The movement of the end of the lever caused by magnetostriction of the specimen causes a displacement of the other end of twice the amount in the opposite direction. Thus, what would have been a displacement of $\Delta L$ is now $-2\Delta L$ by the movable plate of the capacitor. This should be substituted into Eqs. (1)–(4) to obtain the equivalent capacitance change. This is a particularly appropriate way of increasing sensitivity because it does not lead to any significant noise increment during the experiment. The capacitance was measured using AH2550A Ultra- Precession 1 kHz Capacitance Bridge, which utilizes phase sensitivity detection and allows measuring changes in capacitance as small as 0.5 aF. That made our experimental setup more user-friendly, comparing to the bridge-Lock-in amplifier method, which often requires skilled users to tune it up. The described technique can also be applied to ribbons by rolling or piling-up them into bulk samples.[13] To obtain complete information about magnetoelastic properties of the sample the magnetostriction measurements in different directions (with respect to magnetic field) have to be done. Using superconductive split-coil magnet with magnetic field directed along a *c* axis limited us, and a number of others, to measure magnetostriction in the perpendicular mode only. The cell was, therefore, reconstructed to make longitudinal magnetostriction measurements possible [Fig. 2(b)]. Figure 3 presents magnetostriction measurements in MgB2 ceramics along 1.8 mm side length in both longitudinal and transversal modes. The results presented have been analyzed in detail separately.[14,15] Figure 4 presents thermal expansion of the $(La_{0.8}Ba_{0.2})_{0.93}MnO_3$ single crystal measured along different crystallographic directions. The jump in thermal expansion corresponds to the orthorhombic (*Pbmm*) to rhombohedral ($R\bar{3}c$) structural phase transition[16,17].

## C. Thin films

The capacitance technique is also used for investigation of the magnetoelastic behavior in thin films. The idea of capacitance cantilever technique lies in substituting a movable electrode [Fig. 1(a)] by a ferromagnetic cantilever [Fig. 5(a)]. External magnetic field applied in the film plane induces stress, which is wedged firmly to the substrate through magnetoelastic interaction. This stress causes an elastic deformation of the substrate resulting in a deflection of free end of the cantilever [Fig. 5(b)]. The substrate should be thin and sample should be long enough to get reasonable results. The capacitance technique allows measuring the change of capacitance related to the overall curvature of the plate. The analytical solution of this task requires boundary conditions to be considered[18,19]. To determine magnetoelastic contribution to the effective magnetoelastic energy the deflection of film–substrate system should be measured in the perpendicular and parallel mode with respect to the magnetic field [Fig. 5(b)]. The capacitance method was first used by Klokholm[9] for magnetostriction investigation in isotropic films. For single phase films, analytical solution and analytical approach is more complicated and depends on crystalline symmetry of the sample[18–20]. It is easy to show that for small deflections, the changes in the capacitance are connected with the radii $R$ of the cantilever bend as follows:[18]

$$\Delta C = \frac{-C_1^2 L^2}{6\varepsilon\varepsilon_0 SR}, \quad (5)$$

where $C_1$ is the parallel plate capacitance value, $S$ and $L$ the area and length of the film, respectively. The magnetoelastic stress parameter $B^{\gamma,2}$, which is connected with the overall of curvature of the cantilever deflection, can be written[21] as

$$B^{\gamma,2} = (\sigma_\parallel - \sigma_\perp) = \frac{1}{3}\frac{t_s^2}{t_f}\frac{1}{L^2}\frac{Y_s}{(1+\nu)}(\delta_\parallel - \delta_\perp) \quad (6)$$

where $Y_s$ and n $s$ are Young's modulus and Poisson ratio of the substrate, $t_s$ and $t_f$ are the thickness of the substrate and film, respectively. d is the deflection of the bimorph.[22]

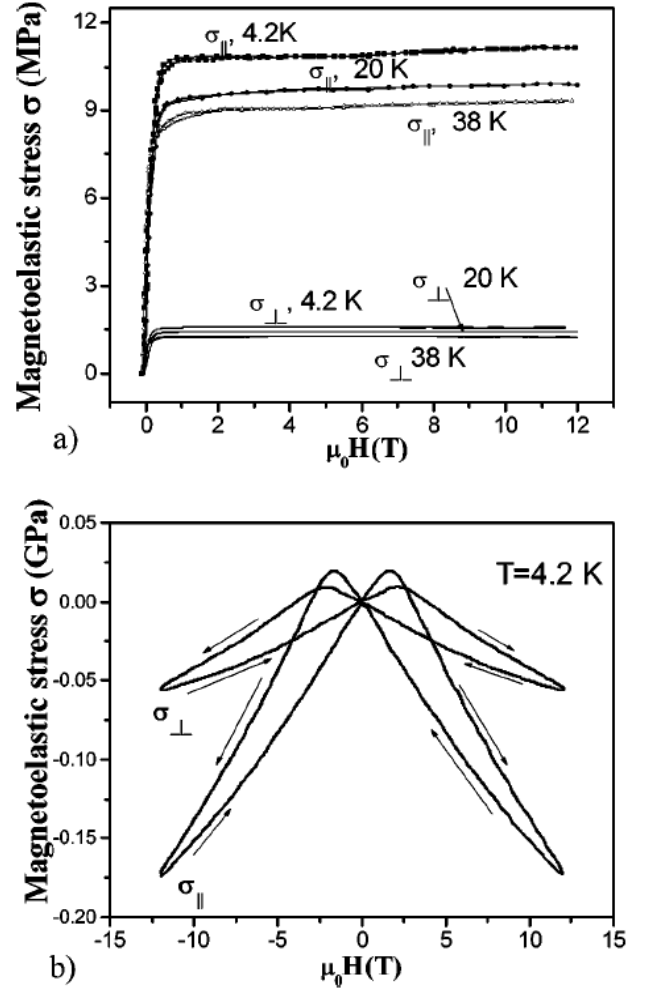

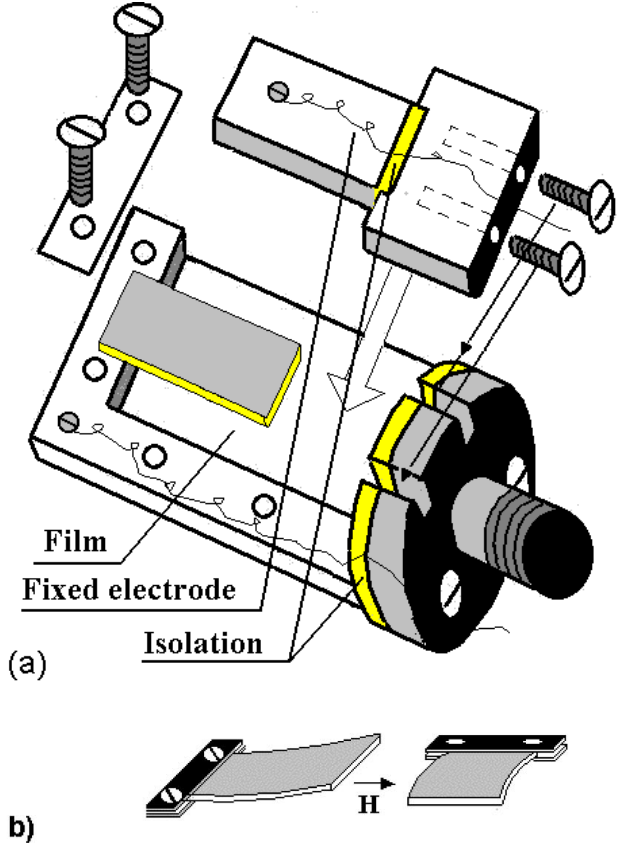

FIG. 5. Capacitance cell adapted for magnetostriction measurements in thin films (a). Schematic of deflection of the magnetoelastic cantilever along $\sigma < 0$ and perpendicular $\sigma < 0$ direction with respect to the applied magnetic field (b).

Figure 6(a) shows magnetoelastic behavior of polycrystalline Ni deposited onto Si(111) substrate. The 3000-Å-thick polycrystalline Ni film was deposited by dc planar magnetron sputtering method with high purity argon (99.9999) as working gas. The Ni film was grown on unheated Si(111) substrate from 3 in. Ni target and with Ar pressure of $5*10^{-3}$ (mbar). The distance from target to substrate was 5 cm. The 400 W output power was used and led to deposition rate close to 50 Å/s. Before the deposition run, the 40 s presputtering of Ni target was made. The Young's modulus of Si substrate was assumed to be temperature independent since elastic constants of Si do not change much with temperature.[23] Observed longitudinal and transversal stresses in Ni/Si(111) cantilever exhibit a positive sign in both cases. That agrees well with the previous room temperature data[24] thus confirming the reliability of our experiment. The measurements were then performed on $La_{0.7}Sr_{0.3}CoO_3$ film deposited onto SAT0.7CAT0.1LA0.2 twins free (001) oriented substrate(SAT- $SrAl_{0.5}Ta_{0.5}O_3$, CAT- $CaAl_{0.5}Ta_{0.5}O_3$, LA-$LaAlO_3$). The 650-Å-thick $La_{0.7}Sr_{0.3}CoO_3$ layer was deposited by pulsed laser deposition without postannealing. The temperature of the substrate was 730 °C. X-ray diffraction investigation showed that the film is single phase, epitaxial, (001) oriented. The sample was cut along appropriate crystallographic directions giving the sample dimensions of 14 mm X4 mm (the longer dimension was along [100] axis). The magnetostriction constant $\lambda_{100}$, is calculated to be 1565 ppm at the magnetic field of 12 T. This is comparable to the magnetostriction of polycrystalline bulk samples.[25] The elastic constant difference $(c_{11}-c_{12})$ was evaluated to be 50 GPa using mechanical properties data.[26]

FIG. 6. Magnetoelastic isotherms of the Ni/Si(111) cantilever (a) and magnetoelastic hysteresis loops of the $La_{0.7}Sr_{0.3}CoO_3$ film deposited onto SAT0.7CAT0.1LA0.2(001) substrate taken at 4.2 K (b).

## III. NOISES

### A. Magnetic fields

In the present work, the three-terminal dilatometer was constructed to be suitable for high magnetic fields (±12T) measurements. The rapid increasing of magnetic field generates spurious voltage in leads and in the shielding, so in our measurements leads were kept short and rigidly tied down since amount of voltage generated by magnetic field in a circuit is proportional to the area that the circuit leads enclose. Moreover, high precision capacitance measurements can be affected by the self inductance of the test cables. Thus, in our experiment, leads were run together and magnetically shielded to minimize induced magnetic voltages. To avoid unwanted parasitic voltage effects, field change was stopped periodically to take a measurement point. In our experiment we have used specially designed AH® low noise coaxial cables optimized for three terminal capacitance measurements. That enables improved stability in the capacitance of the apparatus.

### B. Ground loops

Noise and error voltage can also arise from so-called ground loops. When the source and measuring instrument are both connected to a common ground bus, a loop is formed. A voltage between the source and instrument grounds will initiate a current to flow around the loop. This current will generate an unwanted voltage in parallel with the source voltage. Thus, even small potential difference can cause large currents to circulate and create some unexpected voltage drops. To avoid these ground-loop effects, we have used a single good earth-ground point for the each part of measured system.

## ACKNOWLEDGMENTS

Useful arguments given by Dr. M. Ciria, Professors J. I. Arnaudas and H. Szymczak are gratefully acknowledged. Technical help with substrate preparation by W. Adamczuk is also acknowledged.